\documentclass[prd,twocolumn,preprintnumbers,amsmath,amssymb,nofootinbib]{revtex4}
\usepackage[dvips,final]{graphicx}
\usepackage{amsmath}
\usepackage{amssymb}
\usepackage{url}
\usepackage{textcomp}
\usepackage{graphicx}
\usepackage{bm}
\usepackage{dcolumn}
\usepackage[usenames]{color}

\begin{document}

\title{ Bulk Viscosity and Particle Creation in the Inflationary Cosmology }

\author{Mehdi Eshaghi$^{1,2}$}
\email{eshaghi249@gmail.com}

\author{Nematollah Riazi$^{3}$}
\email{n_riazi@sbu.ac.ir}

\author{ Ahmad Kiasatpour $^{1}$}
\email{akiasat@sci.ui.ac.ir}

\affiliation{ $^1$ Department of Physics , Faculty of Science, University of Isfahan, Isfahan, 81746-73441, Iran}

\affiliation{ $^2$ Astrophysics Sector, SISSA, Via Bonomea 265, I-34136 Trieste, Italy}

\affiliation{$^3$ Department of Physics, Shahid Beheshti University, Tehran 19839, Iran}

 \date{\today} 
  
 \begin{abstract}

We study particle creation in the presence of bulk viscosity of cosmic fluid in the early universe within the framework of open thermodynamical systems. Since the first-order theory of non-equilibrium thermodynamics is non-causal and unstable, we try to solve the bulk viscosity equation of the cosmic fluid with particle creation through the full causal theory. By adopting an appropriate function for particle creation rate of ``Creation of Cold Dark Matter'' model, we obtain analytical solutions which do not suffer from the initial singularity and are in agreement with equivalent solutions of $\Lambda$CDM model. We constrain the free parameter of particle creation in our model based on recent Planck data. It is also found that the inflationary solution is driven by bulk viscosity with or without particle creation.
  \end{abstract}
  
  \maketitle

 \section{Introduction}\label{s.1}
 In the past twenty years, particle creation in cosmological models has drawn the attention of many cosmologists. In some works, particle creation is considered in the context of the thermodynamics of open systems. Within a FRW background, a description for matter creation has been proposed by Prigogine and his coworkers \cite{Prigogine1}. They used the generalized form of the first law of thermodynamics for open systems to describe the flow of energy from the gravitational field to the matter field, resulting in the creation of particles. This is interpreted as an additional negative pressure, which leads to a re-interpretation of the energy-momentum tensor. It means that during an irreversible process, the expanding space-time causes a growth of entropy by producing matter. The negative pressure of particle creation might play the role of dark energy which explains the accelerating universe. Lima and Alcaniz examined the implications of some FRW cosmological models with matter creation against the observations \cite{Lima:1999rt,Alcaniz:1999hu} and later it was shown that these models are consistent with the observational data \cite{Zimdahl:2000zm}.\\
 \indent
 To study the dissipative processes during the evolution of the universe such as entropy production, there is a robust model called ``Dissipative Cosmology'' which is related to scalar dissipation by the space-time symmetries which may be analyzed via the relativistic theory of bulk viscosity \cite{Maartens,Zimdahl:1996ka}.   Dissipative phenomena in a homogenous and isotropic universe are related to bulk viscous pressure. Assuming some dissipative effects of bulk viscosity take place at the cosmological scale, one can explain the accelerating universe \cite{Sen:2000hg}. Indeed, bulk pressures could be the consequence of either of two types of processes. In the First type, different components of the cosmic substratum are coupled due to their different internal equations of state. In this case, cooling of sub-fluids with the expansion of the universe will be different because the system tends to move away from equilibrium. Therefore, it gives rise to a bulk viscous pressure \cite{Weinberg,Schweizer}. In the second type, one can suppose a perfect fluid whose viscous property comes into the picture during the cosmological particle production \cite{Murphy,Hu}. One of the sources of an effective non-vanishing bulk pressure of the cosmic fluid may be the non-conserving particle production which has been studied in some papers \cite{Prigogine1,Zimdahl:1996ka,Calvao,LimaGermano}.\\
 \indent
Some epochs in cosmology, like inflation and reheating (through dissipative processes) are far from equilibrium dynamics. So it is necessary to use relativistic theory of non-equilibrium thermodynamics for them. There are two types of dissipative models based on relativistic non-equilibrium thermodynamics. In the first-order type theory developed by Eckart, Landau and Lifshitz \cite{Eckart,Landau}, authors consider first-order deviation from equilibrium. Consequently, the solutions are non-causal and unstable. In fact, the neglected second-order terms are responsible for this problem. The second-order type is the theory by Israel and Stewart \cite{Israel,Stewart} which removes the above problem by including the neglected terms \cite{Hiscock}. Although the equations of dissipative non-equilibrium thermodynamics are complex, but under a condition called ``isentropic particle production'' which is characterized by constant entropy per particle, one can drive a simple relationship between the viscous pressure and the particle production rate \cite{Zimdahl:1999tn,Chakraborty:2014ora}. Under this condition, the cosmic substratum is not a conventional dissipative fluid but a perfect fluid with varying particle number.\\
\indent
 In the present paper, we will exploit ``isentropic'' condition to overcome complexity of solving the Israel-Stewart second-order equation in the framework of open systems and in the presence of particle creation. This paper is organized as follows. In section \ref{s.2}, we first review viscous FRW universe in the presence of particle creation. Although the results of first-order bulk viscosity model with particle creation are interesting, but as it is non-causal and unstable, it is not a satisfactory theory \cite{Singh:2011dw}. Therefore in section \ref{s.3} we try to solve the Israel-Stewart causal equation under the "isentropic" particle creation condition, exploiting a scenario of ``Creation of Cold Dark Matter'' (CCDM) \cite{Lima:2008qy,Steigman:2008bc,Jesus}. We obtain non-singular analytical solutions for the inflationary and radiation-dominated epochs which are consistent with the equivalent solutions of $\Lambda$CDM model. We show also that the inflationary solution can be a consequence of viscous pressure of cosmic fluid. Finally in section \ref{s.4}, the conclusions of the present work are summarized.

 \section{Viscous FRW Universe with Particle Creation}\label{s.2}

 In ``Dissipative Cosmology'' with particle creation, the Einstein's equation becomes
 \begin{equation}
 R_{\mu \nu  }  -  \frac{1}{2}g_{\mu \nu } R =T_{_{\mu \nu } }^{eff},
 \end{equation}
 where $T_{_{\mu \nu } }^{eff}$ is the most general form of the energy-momentum tensor of cosmic fluid called ``effective energy-momentum tensor''. This tensor can be assumed as
  \begin{equation}
 T_{_{\mu \nu } }^{eff}  = (\rho + p + p_c + \Pi )u_\mu u_\upsilon - (p + p_c + \Pi)g_{\mu \nu } ,
 \end{equation}
 where $\rho$, p, $\Pi$ and $u_\mu$ are the energy density, perfect fluid pressure, bulk viscous pressure and particle-frame 4-velocity, respectively and $p_c$ is defined as the ``creation pressure'' of particles out of the gravitational field \cite{Prigogine1,Calvao,LimaGermano}
 \begin{equation}\label{2.3}
 p_c = - \frac{{(\rho + p)}}{n}\frac{{dN}}{{dV}}
 \end{equation}
where N is the particle number, V is the volume and $n=N/V$ is the particle number density. Also the equation of state of cosmic fluid is supposed as $p = (\gamma-1) \rho$ in which $\,0 \le \gamma  \le 2$ is the ``adiabatic index''. Accordingly, the continuity equation in the presence of bulk viscosity pressure and particle creation pressure takes the form
 \begin{equation}\label{2.4}
 \dot \rho+ 3(\rho +p + p_c +\Pi)H=0,
 \end{equation}
 where the dot denotes derivative with respect to the cosmic time.\\
 \indent
 In the second order theory of the non-equilibrium thermodynamics, the particle 4-flux and entropy 4-flux are defined by $N^{\mu}=nu^{\mu}$ and
 $S^{\mu}=sN^{\mu}-(\tau\Pi^2/2\xi T)u^{\mu}$, in which $u^{\mu}$ is the particle-frame 4-velocity, $s$ is the specific entropy, $\tau\geq0$ the relaxation coefficient for transient bulk viscous effects, $\xi\geq0$ the bulk viscosity coefficient and $T\geq0$ the temperature of the fluid. The transport equation for the bulk viscous pressure is \cite{Israel,Stewart}
 \begin{equation}\label{2.5}
 \Pi+\tau\dot{\Pi}=-\xi\Theta-\frac{1}{2}\tau\Pi\left[\Theta+\frac{\dot{\tau}}{\tau}-\frac{\dot{\xi}}{\xi}-\frac{\dot{T}}{T}\right] ,
 \end{equation}
 where $\Theta=3H$ is the scalar expansion. The functionality of $\xi$, $\tau$ and T are determined according to phenomenological approaches. For example, from the equilibrium thermodynamical equation of state taken from lattice QCD calculations \cite{Karsch:2007jc,Cheng:2007jq} $T=\mu \rho^r$ and $\xi=\varpi\rho+9T_c^4/\omega_0$ with $\gamma=1.318$, $\mu=0.718$, $r=0.213$, $\omega_0=0.5-1.5$ GeV, $T_c\approx 0.19 GeV$ and $\varpi=8.2149\times10^{-5}$.  The viscosity coefficient $\xi$ determines the magnitude of viscous stress relative to expansion in the limit $\tau\rightarrow0$. In this limit we have the first order theory \cite{Eckart,Landau}, i.e. $S^{\mu}=sN^{\mu}$.\\
 \indent 
 The bulk viscous FRW cosmologies are mostly based on the non-causal and unstable first-order thermodynamics. This is the result of the complexity of second-order equations. Supposing the expression in the bracket to be very small, Eq. (\ref{2.5}) reduces to
 \begin{equation}
 \Pi+\tau\dot{\Pi}=-3\xi H .
 \end{equation}
 One way of ensuring that viscous signals do not exceed the speed of light in the truncated theory is to adopt the phenomenological model $\tau=\xi/\rho$ \cite{Maartens}.
 Assuming $\rho \propto \theta^2 $ and $\dot{\Pi}/\theta^3\ll1$, one can find
 \begin{equation}\label{2.7}
 \Pi=-3\xi H.
 \end{equation}
Putting Eqs. (\ref{2.3}) and (\ref{2.7}) and the equation of state into Eq. (\ref{2.4}) and also using $\rho^{\prime}=2\rho H^{\prime}/H$, one can easily obtain Hubble parameter as a function of scale factor \cite{Singh:2011dw}. Although, this result may be interesting but, since it admits dissipative signal with superluminal velocities (it is non-causal) \cite{Israel} and also as all its equilibrium states are unstable \cite{Hiscock} it is therefore not a satisfactory relativistic theory. In the next section we try to solve Israel-Stewart causal equation in the presence of particle creation.\\
 \indent

 \section{Second Order Theory and CCDM}\label{s.3}

 For adiabatic creation, the particle creation rate is given by
 \begin{equation}\label{3.8}
 \Gamma=\frac{{\mathop N\limits^ \cdot  }}{N} = \frac{{\mathop n\limits^ \cdot  }}{n}  +  3\frac{{\mathop R\limits^ \cdot  }}{R} .
 \end{equation}
and therefore Eq. (\ref{2.3}) can be rewritten as
 \begin{equation}\label{3.9}
 p_c = - \frac{{(\rho + p)}}{3H}\Gamma
 \end{equation}

From the Gibbs equation together with Eqs. (\ref{2.4}) and (\ref{3.8}), the variation of the entropy of the cosmic fluid can be written as 
 \begin{equation}\label{3.10}
 nT\dot{s}=-\Pi\Theta-\Gamma(\rho+p) .
 \end{equation}
 Although there is entropy production due to particle creation, but as it was explained in first section, one can get ``isentropic particle production'' with $\dot{s}=0$ and $\Gamma>0$. Therefore, by substituting Einstein's field equations into Eq. (\ref{3.10}) with  $\dot{s}=0$ one obtains \cite{Zimdahl:1996ka,Zimdahl:1999tn,Chakraborty:2014ora}
 \begin{equation}\label{3.11}
 \frac{\Gamma}{\Theta}=1+\frac{2}{3\gamma}\frac{\dot{H}}{H^2} .
 \end{equation}
 
In the present paper, we combine a scenario of ``creation of cold dark matter'' (CCDM) with bulk viscous model. In this CCDM model \cite{Lima:2008qy,Steigman:2008bc,Jesus} which is introduced as an alternative to $\Lambda$CDM model, a new kind of accelerating flat universe with no dark energy which is fully dominated by cold dark matter is studied. The creation rate of these particles is given by \cite{Jesus} 
 \begin{equation}\label{3.12}
 \Gamma=3\alpha\Omega_{dm}^{-1}H
\end{equation} 
where $\alpha$ is a positive constant and $\Omega_{dm}$ is the dark matter density parameter. One can assume that the cold dark matter (CDM) particles are described by a real scalar field maybe as remnant of incomplete reheating of inflaton scalar field after inflation. In this model, the positiveness of the dark matter density at high redshifts and in the matter-dominated era requires that $\alpha\leq\Omega_m$, where $\Omega_m\equiv\Omega_{dm}+\Omega_{baryon}=1$ is the total density parameter in the flat FRW universe. However, we should constraint  the value of $\beta=\alpha\Omega_{dm}^{-1}$ for our model independently. In the following, by using Eqs. (\ref{3.11}) and (\ref{3.12}) we try to solve Eq. (\ref{2.5}).\\
 \indent
Here we adopt a standard power-law form for the temperature $T$ and bulk viscosity coefficient $\xi$ as \cite{Maartens,Coley:1995uh,Lima:1995xz}
 \begin{eqnarray}\label{3.13}
 &&T=\mu \rho^r,\\
 &&\xi=\xi_0\rho^m,
 \end{eqnarray}
where $\mu$, $\xi_0$ and ``m'' are non-negative constants and $r=(\gamma-1)/\gamma$. Using the Friedmann equation $\rho=3H^2$ and Eqs. (\ref{3.13}) and (14), the terms in the bracket of Eq. (\ref{2.5}) can be written as a function of Hubble parameter
 \begin{equation}\label{3.15}
 \frac{\dot{T}}{T}=2r\frac{\dot{H}}{H},\quad  \frac{\dot{\xi}}{\xi}=2m\frac{\dot{H}}{H},\quad  \frac{\dot{\tau}}{\tau}=2(m-1)\frac{\dot{H}}{H} .
 \end{equation}
 On substituting Eq. (\ref{3.15}) and $\dot{\Pi}=\dot{H}\Pi^{`}$ into Eq. (\ref{2.5}) we obtain
 \begin{equation}\label{3.16}
 \Pi^{`}+\Pi\left[\frac{1}{\tau \dot{H}}+\frac{3}{2}\frac{H}{\dot{H}}-(1+r)\frac{1}{H}\right]=-3\frac{\xi}{\tau}\frac{H}{\dot{H}} ,
 \end{equation}
 where $\Pi^{`}$ denotes derivative of $\Pi$ with respect to Hubble parameter. By using Eqs. (\ref{3.11}) and (14), Eq. (\ref{3.16}) can be rewritten as
 \begin{eqnarray}\label{3.17}
 &&\Pi^{`}+\Pi\left[\frac{2}{H\tau \gamma(\Gamma-3H)}+\frac{3}{\gamma(\Gamma-3H)}-(1+r)\frac{1}{H}\right]=\nonumber\\
 &&\frac{-18H^2}{\gamma(\Gamma-3H)}.
 \end{eqnarray}
 Exploiting Eq. (\ref{3.12}) and also the phenomenological model $\tau=\xi/\rho$ \cite{Maartens}, we obtain
 \begin{equation}\label{3.18}
 \Pi^{`}+(kH^{-2m}+lH^{-1})\Pi=-6\frac{H}{\gamma(\beta-1)} ,
 \end{equation}
 where $k=\frac{2}{3^m\xi_0(\beta-1)\gamma}$ and $l=\frac{2\gamma(1-\beta)+\beta}{(\beta-1)\gamma}$.\\
 \indent The analytical solution of Eq. (\ref{3.18}) for a general ``m'' is complicated but in the case of $m=1/2$, it is solvable easily and we obtain
 \begin{eqnarray}\label{3.19}
 \Pi=\frac{-6}{\gamma\beta_{*}(k+l+2)}H^2+C_{1}H^{-(k+l)}
 \end{eqnarray}
 where $\beta_*\equiv\beta-1$ and "$C_{1}$" is a constant of integration. As H is large enough for the inflationary epoch, one can neglect the second term of the above equation and consequently it converts to Eq. (7) in the first order theory. Conversely, in late time that H becomes very small, the second term dominates. Therefore, the second term of Eqs. (\ref{3.19}) shows the difference between first-order and second-order approaches for the bulk viscous pressure.\\
 \indent
 Using Eqs. (\ref{2.4}), (\ref{3.9}) and (\ref{3.19}) we get
 \begin{eqnarray}\label{3.20}
 &&H^{\prime} - \frac{3}{2}\left[\beta_*\frac{\gamma(R)}{R}+\frac{2}{\gamma(R)R\beta_*(k+l+2)}\right]H\nonumber\\
 &&=-\frac{C_1}{3R}H^{-(k+l+1)}.\
 \end{eqnarray}
Here, we use a simple one-parameter function for ``adiabatic index'', $\gamma(R)$ which can describe a transition from an inflationary to a radiation phase \cite{Carvalho}
 \begin{equation}\label{3.21}
 \gamma(R)=\frac{4}{3}\frac{A_*(R/R_{*})^2+(a/2)(R/R_{*})^a}{A_*(R/R_{*})^2+(R/R_{*})^a} ,
 \end{equation}
 where $A_*$ is a constant, $R_{*}$ is a reference value of R and $0\leq a<1$ is a free parameter which has a relationship with the power of the cosmic time t during the inflationary epoch.  
 Substituting Eq. (\ref{3.21}) into Eq. (\ref{3.20}) and integrating, we obtain
 \begin{eqnarray}\label{3.22}
&&H^{2/\delta\beta_*\gamma}\left[A_*(\frac{R}{R_{*}})^2+(\frac{R}{R_{*}})^a\right]^{-4/\delta\gamma}R^{3/\beta_*\gamma}=\nonumber\\
&&-\frac{2C_1}{3\delta\beta_*\gamma}\int dR\left[A_*(\frac{R}{R_{*}})^2+(\frac{R}{R_{*}})^a\right]^{-4/\delta\gamma}R^{(3/\beta_*\gamma)+1}\nonumber\\
&&+C_2 ,
 \end{eqnarray}
 where $\delta=2\sqrt3\xi_0/(2+\sqrt3\xi_0\beta)$ and $C_2$ is a constant of integration.\\
 \indent
 For $R\ll R_*$, $\gamma(R)\rightarrow 2a/3$ and the Hubble parameter is given by
 \begin{equation}\label{3.23}
 H(R)=\left[\frac{2C_1}{12a\beta_*+9\delta}+C_2(\frac{R}{R_{*}})^{6/\delta}R^{9/2\beta_*a}\right]^{\delta\beta_*a/3}.
 \end{equation}
It is noticed that with or without particle creation, the bulk viscous coefficient gives us an inflationary solution and a positive energy density for initial vacuum as $R\rightarrow 0$. In this way, neglecting the terms involving $\beta$ one can obtain a non-singular de-Sitter solution as $R\propto \exp(H_kt)$, where $H_k=(2C_1/9\sqrt{3}\xi_0)^{-\xi_0 a/\sqrt{3}}$. In this picture inflation appears when the viscous stress becomes large enough to create sufficient negative effective pressure. Comparing with the Hubble parameter value in the scalar field inflation, we can find the values of the constants during inflation by using $H_{k}=\pi M_{pl} (rA_s)^{1/2}/\sqrt{2}$ with the primordial scalar amplitude $\ln (10^{10} A_s)=3.089$ \cite{Ade:2013zuv}, where $M_{pl}$ is the Planck mass and $"r"$ is the tensor-to-scalar ratio.\\
 \indent
 For $R\gg R_*$ which leads to $\gamma(R)\rightarrow 4/3$ we find a solution for H
 \begin{equation}\label{3.24}
 H(R)=\left[\frac{2C_1}{24\beta_*+9\delta}+C_2A_*^{3/\delta}(\frac{R}{R_*})^{6/\delta}R^{9/4\beta_*}\right]^{2\delta\beta_*/3}.
 \end{equation}
 On the other hand, the Hubble parameter for the $\Lambda$CDM model in the radiation era (rad) and in the absence of curvature and cosmological constant (which is certainly an acceptable approximation for relativistic high redshift universe) is
 
 \begin{equation}\label{3.25}
 H(R)\simeq 3\times 10^{-23} cm^{-1} (\Omega_{rad}h^2)^{1/2}\left(R/R_0\right)^{-2}.
 \end{equation} 
where $\Omega_{rad}h^2\simeq 4.4\times 10^{-5}$ \cite{Durrer} and $R_0$ is the value of scale factor at the present time. When in the Eq. (\ref{3.24}), the particle creation dominates over bulk viscous coefficient (for large R) we obtain power-law solutions for Hubble parameter as $H\propto R^{4\beta_*}$. In order to constrain the free parameter $\beta$ of the model and have an acceptable power-law form for $H$ in radiation era, we need $0<\beta<1$, i.e. $0<\alpha<\Omega_{dm}$. Considering the recent Planck data for cosmological parameters we find $\alpha<\Omega_{dm}=0.308\pm0.012$ \cite{Planck:2015xua}. Therefore, comparing Eqs. (\ref{3.24}) and (\ref{3.25}) with each other for $\beta\rightarrow 1/2$, i.e. $\alpha\rightarrow 0.154$ one obtains $H\sim R^{-2}$ and consequently putting away the first negligible term in the bracket of Eq. (\ref{3.24}) we find $C_2^{-\delta/3}R_*^2/A_*=1.99\times10^{-25}$.

 Under the condition of ``isentropic particle creation'', the changing rate of thermodynamic parameters is represented by \cite{Zimdahl:1999tn}
 \begin{eqnarray}\label{3.26}
 &&\frac{\dot{n}}{n}=-(\theta-\Gamma),\\
 &&\frac{\dot{T}}{T}=-(\theta-\Gamma)\frac{\partial p}{\partial \rho}.\
 \end{eqnarray}
 So for the selected $\Gamma$, the particle number density and the temperature of the cosmic fluid are given by
 \begin{eqnarray}\label{3.27}
 &&n=n_p \left(\frac{R}{R_*}\right)^{-3(1-\beta)},\\
 &&T=T_p \left(\frac{R}{R_*}\right)^{-3(1-\beta)(\gamma-1)},\
 \end{eqnarray}
 and the particle number can be rewritten as
 \begin{equation}\label{3.28}
 N=N_p \left(\frac{R}{R_*}\right)^{3\beta} ,
 \end{equation}
 where $n_p$, $T_p$ and $N_p$ are some constant references for n, T and N, respectively. For $\beta\rightarrow 0$, $n \propto R^{-3}$ and N would remain constant throughout the time evolution of the universe, i.e. the particle creation rate $\Gamma$ tends to zero. Also we get radiation like solution for $T\propto R^{-1}$ as $\beta\rightarrow 0$ and $\gamma\rightarrow 4/3$.\\
 \indent

 \section{Conclusion}\label{s.4}

 In this work, we studied the role of bulk viscosity in the framework of the particle creation mechanism in flat FRW universe and obtained analytical solutions through the full causal theory of Israel-Stewart. We focused on ``isentropic particle production'' process in which entropy per particle is constant. We solved the transport equation of bulk viscosity for a linear form of $\Gamma(H)$ adopted from CCDM scenario and obtained an expression for the bulk viscous pressure which is different from its counterpart equation in the first-order approach by an extra term that becomes important for very large R and may be responsible for late time acceleration. Using the derived bulk viscous pressure and a varying equation of state parameter $\gamma(R)$, we obtained Hubble parameter, particle number density and the temperature of the cosmic fluid for inflation and radiation eras.\\
\indent
It is interesting that both $\rho$ and H start with a finite value at $t =0$ and end up with a small finite value at very large t. Therefore, in this picture the viscous model offers a desirable possibility for replacing the cosmological constant and explaining the late time acceleration of the universe. Further, It is notable that the inflationary solution can be found with or without particle creation which is due to the constant bulk viscous coefficient. Also for later times, Hubble parameter has power-law behavior which corresponds to the radiation-dominated solutions. We showed that the solutions for the Hubble parameter of both epochs have good consistency with the equivalent solutions of $\Lambda$CDM model based on reported data in recent Planck collaboration papers. Using these data, we constrained the free parameter of particle creation in our model $\alpha<\Omega_{dm}=0.308\pm0.012$.\\
\indent
 It is notable that N is an increasing function of time until $\alpha\sim 0.154$, then it would remain constant forever. Also we found a radiation-like solution for the temperature of the cosmic fluid as $T\propto R^{-1}$ for $\alpha\sim 0.154$ and $\gamma\sim4/3$.
  
\section{Acknowledgement}
    The authors would like to thank Dr. Moslem Zarei and Dr. Shahram Dehdashti for helpful discussions. Mehdi Es'haghi thanks Prof. Carlo Baccigalupi and the organizers of Astrophysics Sector of the International School for Advanced Studies, SISSA for their hospitality during the completion of this work.

\end{document}